\documentclass[12pt]{article}
\usepackage[psamsfonts]{amssymb}
\usepackage{amsmath}
\author{Ehsan Amooghorban\footnote{Ehsan.amooghorban@sci.sku.ac.ir}\,
and Ali Mahdifar\footnote{mahdifar$ _{-} $a@sci.sku.ac.ir} \\ {\small  Department of Physics, Faculty of Basic Sciences, Shahrekord University,
}
\\ {\small P.O. Box 115, Shahrekord 88186-34141, Iran.}}
\title{Quantum dissipative Higgs model}
\begin{document}
\maketitle
\begin{abstract}
By using a continuum of oscillators as a reservoir, we present a
classical and a quantum-mechanical treatment for the Higgs model
in the presence of dissipation. In this base, a fully canonical
approach is used to quantize the damped particle on a spherical
surface under the action of a conservative central force, the
conjugate momentum is defined and the Hamiltonian is derived. The
equations of motion for the canonical variables and in turn the
Langevin equation are obtained. It is shown that the dynamics of
the dissipative Higgs model is not only determined by a projected
susceptibility tensor that obeys the Kramers-Kronig relations and
a noise operator but also the curvature of the spherical space.
Due to the gnomonic projection from the spherical space to the
tangent plane, the projected susceptibility displays anisotropic
character in the tangent plane. To illuminate the effect of
dissipation on the Higgs model, the transition rate between
energy levels of the particle on the sphere is calculated. It is
seen that appreciable probabilities for transition are possible
only if the transition and reservoir's oscillators frequencies to be nearly on resonance.
\end{abstract}

\section{Introduction}
Higgs and Leemon investigated the non-relativistic motion of a
particle embedded on a sphere under the influence of conservative
central potentials \cite{1,2}. These central potentials reduce to
the familiar Coulomb and isotropic oscillator potentials of a
Euclidean geometry when the curvature of the sphere goes to zero.
Higgs defined the motion on a sphere of a constant radius by
means of a gnomonic projection from the motion of a plane,
tangent to the sphere at a given point. The advantage of this
projection over all others for the analysis of particle motion on
a sphere derives from the fact that free particle motion, uniform
motion on a great circle, projects into a rectilinear plane while
its non-uniform motion projects into the tangent plane. In other
words, the projected free particle orbits are the same as those
in the Euclidean geometry: the curvature affects only the speed
of the projected motion. Higgs also showed that this feature
persists in the presence of a central force derived from a
potential V(r), i.e. he proved that the dynamical symmetries in a
sphere are the same as those in the plane. Using the Hamiltonians
as functions of the Casimir operators, the eigenstates and
eigenvalues of the two systems were also obtained in \cite{1} and
\cite{2}, respectively.

In Ref. \cite{4}, based on the Higgs model, we have constructed
the generalized (nonlinear) coherent states \cite{3} of a
two-dimensional harmonic oscillator on a spherical surface to
study some of their quantum optical properties. Accordingly, we
have developed a feasible physical model to generate nonlinear
coherent states on a sphere in a generalized trapped ion system
\cite{5}. The geometry and the algebra of the sphere coherent
states is studied in \cite{6,6.5}. It has been shown that the
structure and properties of sphere-(nonlinear) coherent states
could be explored to studying the curvature effects of physical
space on both transition probability and geometric phase.

Moreover, we have investigated a two-dimensional isotropic
harmonic oscillator on a sphere with a time-dependent radius
\cite{Mirza}. It is seen that variations in the sphere radius
could be represented by a minimally coupled Hamiltonian. As a
realization of the model, a two-dimensional isotropic harmonic
oscillator was considered on a fluctuating background. A simple
golden rule was obtained for the transition probabilities per
unit time between energy levels.

However, the classical and the quantum-mechanical treatment of the
Higgs model have been surveyed so far, are the limit of a more
general formalism for the case where the dissipative effects is
neglected. But, we know that the dissipation is a ubiquitous
phenomenon in real physical systems and in turn leads to the
remarkable effects for more physical systems. We therefore expect
that the dissipative effects affect the dynamical properties of a
quantum particle which is constrained to move on a spherical
surface and lead to some interesting results. The question which
naturally arises in this context, how can we describe the Higgs
model in the presence of the dissipation? The present paper is
intended to respond to this question.

On the other hand, in classical physics the dissipation effects
can be described by introducing a velocity proportional force in
the equation of motion. In quantum mechanics, we can borrow the
idea of force from the classical dynamics. When the force is
conservative, the formulation of the quantum analogue of the
classical motion is straightforward. But, the inclusion of
dissipation by introducing the velocity proportional force
require a time dependent Lagrangian and Hamiltonian and this
leads to the inconsistencies between the equations of motion and
the canonical commutation relations. In addition, the result
depends on our choice of the Lagrangian or the Hamiltonian.
Several approaches have been developed to circumvent this
problem~\cite{Ford1965}-\cite{Weiss2008}. The most successful and
rather general approach is based on a reservoir concept by
introducing it explicitly in the quantization process. Although,
knowing the microscopic details of the reservoir is not
necessarily needed.

Recently, we have presented a Lagrangian scheme to the
quantization of the dissipative systems~\cite{Amooghorban2014}.
This approach allows us to obtain a deeper understanding from the
mechanism of the dissipation in the higgs model, as well as it
brings forth the simplest way in which the dissipation and the
fluctuation effects emerge from the classical to the quantum
relativistic domain. In the quantization process, it is seen that
a reservoir containing of a continuum of three dimensional
harmonic oscillators mimics the isotropic~\cite{Kheirandish2010}
and anisotropic dissipative materials~\cite{Amooghorban2014} and
even the nonlinear dissipative materials~\cite{Amooshahi2010}.
Consequently, the main system including the reservoir is
described by a time-independent Lagrangian and Hamiltonian and
the quantization of the dissipative system is performed in a
completely standard fashion.

The paper is organized as follows. In Sec. 2, we propose a
Lagrangian for the total system and investigate a classical
treatment of a dissipative particle embedded on a spherical space
by means of the gnomonic coordinate. In Sec. 3, we use the
Lagrangian introduced in the Sec. 2 to canonically quantize the
system and obtain the corresponding Langevin equation.
Subsequently, in Sec. 4, as a simple application, we calculate
the transition rate between energy levels of the particle on the
sphere and show that how the curvature of physical space and the
dissipative effect compete in the transition rate of
aforementioned particle. Finally, the summary and concluding
remarks are given in section 5.
\section{Classical dynamics}
The dissipative effects for more physical systems are important
and can not be ignored. One of these interesting systems, which
has been studied by Higgs~\cite{1}, is a quantum particle on a
two-dimensional sphere under the action of a conservative central
force. In the higgs model, we may imagine that the dissipation
effects introduce to our formalism by interacting the main system
with the quantum vacuum field and/or a lossy medium. This
approach based on the Lagrangian quantization scheme therefore
prepares the grounds to extend the Higgs model to include the
dissipative effects. In~\cite{Amooghorban2014}, the classical and
the quantum description of a dissipative system was accomplished
by modeling a system interacting with a reservoir. The reservoir
was assumed there made up a continuum of three dimensional
harmonic oscillators labeled by a continuous parameter $\omega$.
This reservoir modeling in fact was inspired on the basis of the
microscopic Hayfield model of a dielectric~\cite{Hopfield1958}.
We apply this approach to examine the classical and the quantum
treatment of the dissipative particle embedded on the spherical
space. Let us start the following classical Lagrangian for the
total system
\begin{equation} \label{total Lagrangian}
L(t) = L_{\rm r}(t)  + L_{\rm s}(t) +L_{\rm int}(t).
\end{equation}
The first term $L_{\rm r}(t)$ is the Lagrangian of the reservoir
which is a continuum of three dimensional harmonic oscillators
defined by
\begin{eqnarray} \label{Lagrangian of reservoir}
L_{\rm r}(t) \, = \frac{1}{2}\int_0^\infty  d\omega\, \left[{{
\dot X}_i^2(\omega,t)- \omega ^2 { X}_i^2(\omega,t)}\right],
\end{eqnarray}
where ${ X}_i(\omega,t),\, i=1,2,3$, are the Cartesian components
of the reservoir variable and the Einstein convention  for
repeated indices is used.\\

The second term $L_{\rm s}(t)$ is the Lagrangian of our main
system on a sphere with radius $R$ under the influence of a
conservative centeral force derived from a potential $V(|{\bf
q}|)$, given by
\begin{equation}\label{Lagrangian for main system}
L_{\rm s}(t) \, = \frac{1}{2}m\,{\bf \dot q}^2(t) -V(|\bf q|),
\end{equation}
where the Cartesian coordinates of the particle embedded on the
spherical space are designated by ${\bf q}=(q_{1}, q_{2}, q_{3})$
and is assumed that they satisfy the sphere equation:
$q_{1}^{2}+q_{2}^{2}+q_{3}^{2}=\frac{1}{R^{2}}$.

In~\cite{1,2}, the authors introduced a gnomonic projection by means
of the projection of a point on the sphere through its center
onto the tangent plane in the embedding space. The advantage of
this projection over all others for the analysis of the motion of
a particle on a sphere stems from the fact that a free particle
on a sphere just affects by the curvature of the sphere and the
form of the central potentials does not change by the curvature
of the sphere. We can use this advantage by applying the gnomonic
projection to the Lagrangian of our system. To this end, we first
express the relationship between the Cartesian coordinates ${\bf
q}$ of the particle embedded on the sphere and two Cartesian
coordinates $x_{\alpha} (\alpha=1,2)$ of the tangent plane as
\begin{equation}\label{relation between two coordinates}
{\bf q}=\Lambda^{-1}(x_{1},x_{2},\lambda^{-\frac{1}{2}}),
\end{equation}
where $\lambda=R^{-2}$ is the curvature of the sphere,
$\Lambda=\sqrt{1+\lambda r^2}$ and $r^2=x_\beta x_\beta$. On the
other hand, for further calculations, it is convenient to obtain
the metric tensor of the physical space. According to above
relation, It can be easily shown that the metric tensor of
physical space in terms of the tangent space coordinates is
$\bar{\bar{g}}({\bf x},\lambda)=\Lambda^{-2}({\bf
1}-\lambda\Lambda^{-2}{\bf x } {\bf x})$. This enable us to
rewrite the Lagrangian (\ref{Lagrangian for main system}) in term
of the new variable $x_{\alpha}$ as
\begin{equation}\label{Lagrangian for main system in tangent plane}
L_{\rm s} (t)\, =\frac{1}{2\Lambda^2}\left({{\bf
\dot{x}}^2-\lambda\frac{({\bf x}\cdot {\bf
\dot{x}})^2}{\Lambda^2} }\right)-V({\bf x}).
\end{equation}

The Lagrangian $ L_{\rm int}(t) $ is the interaction term defined
by
\begin{eqnarray} \label{Lagrangian of interaction}
L_{\rm int}(t)  = \int_0^\infty {d\omega \,}f(\omega )\,\dot {
q}_i (t)\,{X}_{i}(\omega,t).
\end{eqnarray}
Similar to the main system Lagrangian ~(\ref{Lagrangian for main
system in tangent plane}), we can rewrite $ L_{\rm int}(t) $ in
term of the coordinate of the tangent plane as
\begin{eqnarray}\label{Lagrangian of interaction in tangent plane}
L_{\rm int}(t)  = \int_0^\infty {d\omega \,}f(\omega
)\,\bar{\bar{a}}_{i \beta }({\bf x},\lambda) \, \dot {x}_\beta
\,{X}_{i}(\omega,t),
\end{eqnarray}
where the geometric tensor $\bar{\bar{a}}({\bf x},\lambda)$ is
defined in term of the metric tensor as $\bar{\bar{a}}_{i \alpha
}\bar{\bar{a}}_{i \beta}({\bf x},\lambda)=\bar{\bar{g}}_{\alpha
\beta}({\bf x},\lambda)$. Now, the total Lagrangian~(\ref{total
Lagrangian}) has been specified in term of the coordinates of the
tangent plane, we can use the Euler-Lagrange equation to obtain
the classical equations of motion. We find that the classical
equations for dynamical variables ${\bf x}$ of the spherical
particle and ${\bf X}(\omega,t)$ of the reservoir, are
respectively
\begin{eqnarray}\label{classical equation x}
{\ddot x}_\alpha -\frac{2\lambda({\bf x}\cdot\dot{{\bf x}})}{\Lambda^2}\dot{x}_\alpha
+\frac{\Lambda^4 }{r} \frac{d V(r)}{dr} x_\alpha = -\bar{\bar{g}}^{-1}_{\alpha \beta}{\dot  { R}}_\beta ,
\end{eqnarray}
and
\begin{eqnarray}\label{classical equation X}
\ddot {X}_i(\omega,t) + \omega ^2\, X_i(\omega,t) =\bar{\bar{a}}_{i \beta }({\bf x},\lambda)\,\dot{x}_\beta (t)f(\omega ),
\end{eqnarray}
where $ \bar{\bar{g}}^{-1}({\bf x},\lambda)= \Lambda^2\left({
{\bf 1} +\lambda{\bf x} {\bf x} }\right)$ is the inverse of the
metric tensor $\bar{\bar{g}}$, and the components of the field
${\bf R}$ are defined by
\begin{equation}\label{definition R}
R_\alpha({\bf x},\lambda; t) = \bar{\bar{a}}_{j \alpha }({\bf
x},\lambda) \int_0^\infty{d\omega } f (\omega )X_j(\omega,t).
\end{equation}
The formal solution of the field equation (\ref{classical
equation X}) is obtained as
\begin{eqnarray} \label{solution of equation X}
{ X}_i (\omega,t )   &=&  { \dot{X}}_i (\omega,0 ) \frac{{\sin \omega t}}{\omega }
+  { X_i} (\omega,0 )\cos \omega t \nonumber\\
&&+ \bar{\bar{a}}_{i \beta }({\bf x},\lambda)\int_0^t  {dt'}
\frac{{\sin \omega (t - t')}}{\omega }  { f}(\omega) { \dot x}_\beta(t^\prime)
\end{eqnarray}
where the two first term is associated with the homogeneous
solution of Eq.~(\ref{classical equation X}) and the third term is
related to the inhomogeneous one. By substituting the solution ${
X}_i (\omega,t ) $ into the Eq.~(\ref{definition R}), the field
$\bf R$ is obtained as
\begin{equation}\label{R field}
{{ R}_\alpha}({\bf x},\lambda; t) = \int_0^\infty  {dt'\,}
\bar{\bar{\chi}} _{\alpha \beta}({\bf x},\lambda; t-t'){\dot {
x}}_\beta(t')\, + { R}_{\alpha}^{\rm N}({\bf x},\lambda; t).
\end{equation}
The medium susceptibility tensor projected on the tangent space,
$\bar{\bar{\chi}} _{\alpha \beta}({\bf x},\lambda; t)$, is given
by
\begin{eqnarray} \label{susceptibility tensor}
\bar{\bar{\chi}} _{\alpha \beta}({\bf x},\lambda; t) = \bar{\bar{g}}_{\alpha \beta}({\bf x},\lambda)\gamma(t),
\end{eqnarray}
where $\gamma(t)=\int_0^\infty d\omega\, f(\omega)\, \frac{\sin
\omega t}{\omega}{\rm \Theta(t)}$ is the medium susceptibility
function felt by the particle on the sphere. Incidentally, this
relation shows that the susceptibility function $\gamma$ and in
turn the projected susceptibility tensor $\bar{\bar{\chi}}$
satisfy the significant Kramers-Kronig relations, as they must be.
Mathematically, these two susceptibility tensors are related to
each other via the metric tensor. Therefore, we except that, even
though the medium susceptibility function $\gamma(t)$ is
isotropic and homogenous for the particle on the sphere, the
projected susceptibility tensor on the tangent plane,
$\bar{\bar{\chi}}$, is not only anisotropic and inhomogeneous but
also depends on the curvature of the sphere. This is not
surprising, because even in the absence of the dissipative
effects, the free particle motion on a sphere is projected onto
rectilinear, but non-uniform motion on the tangent plane~\cite{1}.

Given an arbitrary susceptibility function $\gamma$, the coupling
tensor $f(\omega)$ can be uniquely determined. By taking the
inversion of Eq.~(\ref{susceptibility tensor}), the corresponding
coupling function is obtained as following:
\begin{eqnarray}\label{definition coupling function}
f(\omega ) &=& \sqrt{\frac{{2\omega }}{\pi
}\, \bar{\bar{g}}^{-1}_{\alpha \beta}({\bf x},\lambda)\,{\rm Im}[{\bar{\bar{\chi}}}_{\alpha \beta}({\bf x},\lambda;\omega)]}\nonumber\\
&=&\sqrt{\frac{{2\omega }}{\pi}\, {\rm Im}[\gamma(\omega)]},
\end{eqnarray}
where ${\gamma} (\omega)$ is defined as
\begin{eqnarray}\label{definition fourier integral for susceptibility}
{\gamma} (\omega) &= &\int_{0}^\infty {dt
}\, {\gamma }(t)\, e^{\imath\omega t}\nonumber\\
&=&\int_{ 0}^\infty d\omega' \frac{f^2(\omega )}{\omega'^2-\omega^2+\imath 0^+}.
\end{eqnarray}

The second term in Eq.~(\ref{R field}) are components  of a noise
vector which is defined in terms of the medium susceptibility
function $\gamma(t)$ and the geometry tensor $\bar{\bar{a}}({\bf
x},\lambda)$ as
\begin{eqnarray} \label{R noise}
{R}_{\alpha}^{\rm N}({\bf x},\lambda; t) &=&
\bar{\bar{a}}_{\alpha j }({\bf x},\lambda)\int_0^\infty  {d\omega } \,\sqrt{\frac{{2\omega }}{\pi}\, {\rm Im}[\gamma(\omega)]}\,\nonumber\\
&&\times\left(
{{\dot{X}}_j(\omega,0)\frac{{\sin \omega t}}{\omega } + { X}_j(\omega,0)
\cos \omega t} \right).
\end{eqnarray}
In this manner, the field ${{ \mathbf{R}}}({\bf x},\lambda; t)$ is
completely determined in term of the medium susceptibility
function $\gamma(t)$, the geometry tensor $\bar{\bar{a}}({\bf
x},\lambda)$ and the noise vector ${\bf R}^{\rm N}({\bf
x},\lambda; t)$. Now, by inserting Eq.~(\ref{R field}) into
Eq.~(\ref{classical equation X}), we obtain the classical
Langevin equation as
\begin{eqnarray} \label{classical Langevin equation}
{\ddot x}_\alpha -\frac{2\lambda({\bf x}\cdot\dot{{\bf x}})}{\Lambda^2}\dot{x}_\alpha+\, \int_0^\infty  {dt'\,} \dot{\gamma}( t-t')\, {\dot {
x}}_\alpha(t')+\frac{\Lambda^4 }{r} \frac{d V(r)}{dr} x_\alpha = {\xi}^{\rm N}({\bf x},\lambda;t),
\end{eqnarray}
where the stochastic force induced by the reservoir, ${\xi}^{\rm
N}({\bf x},\lambda;t)$, is given by
\begin{eqnarray}
{\xi}^{\rm N}({\bf x},\lambda;t)=-\bar{\bar{g}}^{-1}_{\alpha
\beta}({\bf x},\lambda){\dot  { R}}^{\rm N}_\beta({\bf
x},\lambda;t).
\end{eqnarray}
It is clear from Eq.~(\ref{classical Langevin equation}) that
the dissipative effects appear as a frictional force comprising
the linear functional of the history of the velocity of the
particle on the sphere and the inhomogeneous stochastic force
${\xi}^{\rm N}({\bf x},\lambda;t)$. This conclusion is expected
by referring to the consideration of Ref.~\cite{Amooghorban2014}, of course with
a difference that the metric and the curvature of the sphere are
arrived to our computation. In the limit $\gamma \rightarrow 0$,
the classical Langevin equation~(\ref{classical Langevin
equation}) smoothly tend to the non-dissipative one which is
indeed consistent with the result reported in~\cite{1}.
%
%
\section{Quantum dynamics of the dissipative Higgs model}
By using the Lagrangian (\ref{total Lagrangian}), we can now
calculate canonical conjugate momentum associated with the
dynamical variables ${\bf x}(t)$ and ${\bf X}(\omega,t)$ ,
respectively, as
\begin{eqnarray} \label{canonical momentum}
&&  {\bf p} (t) = \frac{\partial L}{\partial \,\dot {\bf x} }
= \Lambda^{-2}\dot {\bf x} -\Lambda^{-4}({\bf x} \cdot \dot{{\bf x}}) {\dot{\bf x}}+ {\bf R},\\
&& {\bf P}(\omega,t)  = \frac{\delta L}{\delta \dot {\bf
X}(\omega,t)} = \dot {\bf X}(\omega,t).
\end{eqnarray}
To consider the quantum dissipative Higgs model, we impose
equal-time commutation relations among the variables and the
corresponding conjugates momentum as
\begin{eqnarray} \label{commutation relation x}
 &&[x_\alpha(t), p_\beta(t)] = \iota \hbar\, \delta _{\alpha \beta},
\end{eqnarray}
\begin{equation}\label{commutation relation X}
[X_i(\omega,t), P_j(\omega',t)] = \iota \hbar\, \delta _{ij}\,
\delta (\omega  - \omega ').
\end{equation}
Therefore, by using the Lagrangian (\ref{total Lagrangian}) and
the relations for the canonical conjugate variables
in~(\ref{commutation relation x}) and~(\ref{commutation relation
X}), we obtain the quantum Hamiltonian of the total system as
\begin{eqnarray} \label{total Hamiltonian}
&&{\cal H}= {\cal H}_{\rm Higgs}({\bf x},{\bf p} - {\bf
R};\lambda)+\frac{1}{2} \int_0^\infty d\omega \left({{\bf P}^2
(\omega,t)+ \omega ^2 {\bf X}^2(\omega,t)}\right),
\end{eqnarray}
where
\begin{eqnarray} \label{Hamiltonian H0}
{\cal H}_{\rm Higgs}({\bf x},{\bf p} ;\lambda)=\frac{1}{2}(\pi_\alpha \pi_\alpha+ \frac{\lambda}{2}L_{\alpha \beta}L_{\alpha \beta})+V(\bf x),
\end{eqnarray}
and ${\boldsymbol \pi}={\bf p}+\frac{\lambda}{2}\left[{{\bf
x}({\bf x}\cdot{\bf p})+({\bf x} \cdot{\bf p}){\bf x}}\right]$.
The above Hamiltonians~(\ref{total Hamiltonian})-(\ref{Hamiltonian
H0}), together with the commutation relations ~(\ref{commutation
relation x})-(\ref{commutation relation X}), complete our
procedure to describe the dissipative Higgs model as quantum
mechanically. It is important to note that the Hamiltonian of the
main system ${\cal H}_{\rm Higgs}$ has the same form of the Higgs
Hamiltonian~\cite{1}, but with a difference that the momentum
variable ${\bf p}$ is replaced by the expression $({\bf p} - {\bf
R})$. Indeed, this result justify the minimal coupling scheme was
presented in~\cite{Amooghorban2014} to introduce the dissipation
effects into our formalism.

To facilitate our calculations, let us introduce the annihilation
operator
\begin{eqnarray} \label{annihilation operator b}
&& b_i (\omega ,t) =\frac{1}{ \sqrt {2\hbar\omega}} [\omega X _{
i }(\omega,t)+ \imath P_{i}(\omega,t)].
\end{eqnarray}
By using the canonical commutation relations~(\ref{commutation
relation x}) and~(\ref{commutation relation x}), we readily obtain
\begin{eqnarray} \label{commutation relation b}
&& \left[ {b_i  (\omega ,t),b_{j}^ \dag (\omega ',t)} \right] = \delta _{ij}\, \delta (\omega  - \omega '),
\end{eqnarray}
with all other equal-time commutators being zero. By inverting
Eq.~(\ref{annihilation operator b}), we can express the canonical
conjugate variables  $ X_i (\omega,t)$ and $ P_i (\omega,t)$ in
terms of the creation and annihilation  $b_i ^ \dag (\omega ,t)$
and $b_i (\omega ,t) $. By inserting these relations into the
Hamiltonian~(\ref{total Hamiltonian}), we obtain the Hamiltonian
of the total system as follows:
\begin{eqnarray} \label{total Hamiltonian in term of b}
{\cal H} = {\cal H}_{\rm Higgs}({\bf x},{\bf p} - {\bf R};\lambda)+ {\cal H}_{ r},
\end{eqnarray}
where
\begin{equation}\label{R noise in term of b}
R_\alpha({\bf x},\lambda) = \bar{\bar{a}}_{j \alpha }({\bf x},\lambda)
{\int_0^\infty {d\omega }}\,\sqrt{\frac{{\hbar }}{\pi}\, {\rm Im}[\gamma(\omega)]}\left[ {b_j (\omega
,t) + b_j ^ \dag ( \omega ,t)} \right],
\end{equation}
and
\begin{eqnarray} \label{Hamiltonian for reservoir}
&& {\cal H}_{r} = :{\int {d\omega } } \,\hbar \omega \,\,b_j ^ \dag  (\omega ,t)\,b_j  (\omega ,t):
\end{eqnarray}
is the Hamiltonian of the reservoir in normal ordering form. To
study the dynamics of the dissipative particle on the sphere under
the central potential $V(\bf q)$, let us proceed in the Heisenberg
picture. By using commutation relations~(\ref{commutation
relation x}),~(\ref{commutation relation X}) and the total
Hamiltonian~(\ref{total Hamiltonian}), the equations of motion
for the canonical variables ${\bf X }$ and ${\bf P}$ are
obtained, respectively, as
\begin{equation}\label{Heisenberg equation for X}
\dot {X}_i(\omega,t) = \frac{\iota }{\hbar }[{\cal H},
X_i(\omega,t)] = P_i(\omega,t),
\end{equation}
\begin{equation} \label{Heisenberg equation for Q}
\dot {P}_i(\omega,t)= \frac{\iota }{\hbar }[{\cal H},
P_i(\omega,t)] = - \omega ^2 X_i(\omega,t) + f(\omega ){
\bar{\bar{ a}}}_{i \beta} \,{\dot{x}}_\beta  .
\end{equation}
It can be easily shown that the combination of these equations
are identical to the classical equation~(\ref{classical equation
X}) with the formal solution~(\ref{solution of equation X}).
Calculations analogous to those of~(\ref{Heisenberg equation for
X}) and~(\ref{Heisenberg equation for Q}) give the following
Heisenberg equations for the dynamical variables ${\bf x}$ and
${\bf p}$, respectively, as
\begin{eqnarray} \label{Heisenberg equation for x}
&& {\bf \dot x} = \frac{\iota }{\hbar }\left[ {{\cal H},{\bf q}(t)}
\right] = \Lambda^2({\bf 1}+\lambda {\bf x} {\bf x}) \cdot ({\bf p} - {\bf R}) ,
\end{eqnarray}
\begin{eqnarray} \label{Heisenberg equation for p}
&& {\bf \dot p} = \frac{\iota }{\hbar }\left[ {{\cal H},{\bf p}(t)}
\right] =  -\frac{\lambda\left({{\bf x}\dot{x}^2+\dot{x}({\bf x}\cdot\dot{\bf x})}\right)}{\Lambda^4}+\frac{2\lambda^2{\bf x}({\bf x}\cdot\dot{\bf x})^2}{\Lambda^6}- {\nabla} V ({\bf x}).
\end{eqnarray}
By combining the recent Heisenberg equations and eliminating the reservoir
degrees of freedom by inserting the solution~(\ref{solution of equation X}) into Eq.~(\ref{Heisenberg equation for x}), we obtain the
quantum analogous of the Langevin equation~(\ref {classical
Langevin equation}) as
\begin{eqnarray} \label{quantum Langevin equation}
{\ddot x}_\alpha -\frac{2\lambda({\bf x}\cdot\dot{{\bf x}})}{\Lambda^2}\dot{x}_\alpha+\, \int_0^\infty  {dt'\,} \dot{\gamma}( t-t')\, {\dot {
x}}_\alpha(t')+\frac{\Lambda^4 }{r} \frac{d V(r)}{dr} x_\alpha = {\xi}^{\rm N}({\bf x},\lambda;t),
\end{eqnarray}
where the explicit form of the noise operator ${\xi}^{\rm N}({\bf
x},\lambda;t)$ is written in terms of the bosonic
operator~(\ref{annihilation operator b}) as
\begin{equation} \label{noise operator}
{\xi}_{i}^{\rm N}({\bf x},\lambda;t) =
- \bar{\bar{g}}^{-1}_{\alpha \beta}\int_0^\infty  {d\omega } \,\sqrt{\frac{\hbar \,{\rm Im}[\gamma (\omega)]}{\pi}}\,
\left({ (\dot{\bar{\bar{a}}}_{j \beta}-\imath \omega \bar{\bar{a}}_{j \beta} )
{b}_j(\omega,0)e^{-\imath \omega t} + {\rm h.c.}} \right).
\end{equation}
From these results, the close formal analogy between the classical
and quantum treatments of the dissipative Higgs model can be seen.
We can easily show that in the classical limit, $\hbar \rightarrow
0$, Eq.~(\ref{quantum Langevin equation}) properly reduces to the
classical analogue one~(\ref{classical Langevin equation}), in
the sense that in this limit all commutators vanish and the
equation of motion becomes an equation for c-numbers driven by a
noise force function.
%
\subsection{Transition rates}
In this section, as an application of the presented approach, we
investigate briefly the question of transition probabilities per
unit time between two arbitrary states. For this purpose, we
consider only the first order processes and write the
Hamiltonian~(\ref{total Hamiltonian}), up to first order of
approximation in $\lambda$, as
\begin{equation}\label{approximated Hamiltonian}
{\cal H} = {\cal H}_0  + {\cal H}_{\rm int} ,
\end{equation}
where
\begin{equation}\label{H_0}
{\cal H}_0  = {\cal H}_{\rm Higgs} ({\bf x},{\bf p} ;\lambda) + {\cal H}_{\rm r} , \\
\end{equation}
and
\begin{eqnarray}\label{H_int}
{\cal H}_{\rm int}  =   \int_0^\infty  {d\omega } \,\sqrt{\frac{\hbar \,{\rm Im}[\gamma (\omega)]}{\pi}}\,
\{ ({ p}_\beta {\bar{\bar{a}}}_{j \beta}+\lambda { p}_\beta \bar{\bar{a}}_{j \beta} { x}_\alpha{ x}_\alpha)
{b}_j(\omega,t) + {\rm h.c.} \}.
\end{eqnarray}
In the Hamiltonian~(\ref{approximated Hamiltonian}), the term
$R^2$ has been omitted since it can give rise to the transitions
only in second or higher orders while we study here the first
order processes in which the number of photons changes by $\pm1$.
In addition, for simplicity, we restrict our concentration to a
sufficiently small curvature. Therefore, the terms including the
second and higher power of the curvature $\lambda$ have been
ignored too. We calculate the transition probability at time $t$
from an initial state $|m,\,M_{j}(\omega)\rangle $ of ${\cal
H}_0$ to a final state $ | n,\,M_{j}(\omega)+1\rangle$ by
treating ${\cal H}_{\rm int} $ as a perturbation,
\begin{eqnarray}\label{transition probabilities formula}
P_{m,\,M_{j}(\omega)\rightarrow n\neq m,\,M_{j}(\omega)-1}
= \nonumber\\
&& \hspace{-2cm}\frac{1}{\hbar^2}\left|{ \int_{0}^{t} dt'\,
\langle  n,\,M_{j}(\omega)-1 | e^{\frac{\imath}{\hbar}{\cal H}_0 t'} \, {\cal H}_{\rm int} \,
 e^{-\frac{\imath}{\hbar}{\cal H}_0 t'} | m,\,M_{j}(\omega) \rangle }\right|^2  \nonumber\\
&&\hspace{-2cm}=  \frac{1}{ 8 \hbar \omega}\left|{ \int_{0}^{t} dt'\, f(\omega)\sqrt{M_{j}(\omega)}
e^{\imath (\omega_{nm}-\omega)t'}  ( V_{j,nm} + \lambda {V'}_{j,nm})}\right|^2,\nonumber\\
\end{eqnarray}
where $\omega_{nm}=(E_n-E_m)/\hbar$,
\begin{eqnarray}
V_{j,nm}= \langle  n| ( { p}_\beta {\bar{\bar{a}}}_{j \beta} +
{\rm h.c.} ) | m \rangle ,
\end{eqnarray}
and
\begin{eqnarray}
{V'}_{j,nm} = \langle  n| ( { p}_\beta \bar{\bar{a}}_{j \beta} {
x}_\alpha{ x}_\alpha + {\rm h.c.} ) | m \rangle.
\end{eqnarray}
If $V_{j,nm}$ and ${V'}_{j,nm}$ are explicitly time-independent,
that is the case here, we may carry out the integral of Eq.
(\ref{transition probabilities formula}) and obtain
\begin{eqnarray}\label{transition probabilities M-1}
P_{m,\,M_{j}(\omega)\rightarrow n\neq m,\,M_{j}(\omega)-1}
&=&\frac{{\rm Im}[\gamma (\omega)]M_{j}(\omega)\sin^2[(\omega_{nm}-\omega)t/2]}{4 \pi \hbar[(\omega_{nm}-\omega)t/2]^{2}}\nonumber\\
&& \times \mid V_{j,nm} + \lambda {V'}_{j,nm}\mid^2.
\end{eqnarray}
As is known, the factor
$\frac{\sin^2[(\omega_{nm}-\omega)t/2]}{[(\omega_{nm}-\omega)t/2]^{2}}$
in Eq. ~(\ref{transition probabilities M-1}) is a very strongly
peaked function of $\omega_{nm}$. At $\omega_{nm}=\omega$ its
amplitude increases as $t$ and decreases to zero when
$\omega_{nm}-\omega=2\pi/t$. The probability that $V_{j,nm}$ and
${V'}_{j,nm}$ induce the main system to make a transition between
state $| n \rangle $ and $| m \rangle $ is thus very small unless
energy is conserved between the initial and final states. We may
therefore replace this highly packed factor by the Dirac function
$2 \pi t\, \delta(\omega_{nm}-\omega)$ in Eq. ~(\ref{transition
probabilities M-1}). Thus, the transition probability per unit
time for the absorption of a phonon is given by
\begin{eqnarray}\label{Gamma M-1}
\Gamma_{m,\,M_{j}(\omega)\rightarrow n\neq m,\,M_{j}(\omega)-1}
=\frac{{\rm Im}[\gamma (\omega)]}{2 \hbar}  M_{j}(\omega) \mid
V_{j,nm} + \lambda {V'}_{j,nm}\mid^2 \delta(\omega_{nm}-\omega),
\end{eqnarray}
which is the desired golden rule.

In a similar way, the transition probability per unit time from
an initial state $|m,\,M_{j}(\omega)\rangle $ of ${\cal H}_0$ to
a final state $ | n,\,M_{j}(\omega)+1\rangle$ is obtained as
\begin{equation}\label{Gamma M}
\Gamma_{m,\,M_{j}(\omega)\rightarrow n\neq m,\,M_{j}(\omega)+1}
=\frac{{\rm Im}[\gamma (\omega)]}{2 \hbar}  [M_{j}(\omega)+1]
\mid V_{j,nm} + \lambda {V'}_{j,nm}\mid^2
\delta(\omega_{nm}+\omega).
\end{equation}
This equation is the probability of the main system to decay
spontaneously when no radiation was initially present.

To summarize, we obtain the appreciable transition probability
only if $E_{n}\simeq E_{m}-\hbar \omega$ (stimulated emission) or
$E_{n}\simeq E_{m}+\hbar \omega$ (absorption). The particle on the
sphere may, therefore, be changed from lower (upper) states to
upper (lower) ones by using suitable frequencies of the
reservoir's oscillators.

\section{Summary and concluding remarks}
In this paper, based on the Lagrangian scheme, the classical and
the quantum-mechanic treatment of the damped particle on a sphere
under the action of a conservative central force is investigated.
In this approach, the dissipation and the fluctuation effects are
introduced to our formalism by interacting the reservoir
containing of continuum of three dimensional harmonic oscillators
with the main system. As a direct consequence of this
interaction, the explicit expressions for the projected
susceptibility and quantum noise in terms of the coupling
function and the metric of the physical system are obtained. It
is shown that for the damped particle in the tangent plane,
though the reservoir is assumed to be isotropic on the spherical
surface, the projected susceptibility displays anisotropic
character.

In Heisenberg picture, the Hamiltonian and in turn the quantum Langevin equation is derived. Of course,
this Hamiltonian appears as a minimal coupling form somehow that the conjugate
momentum of the particle coupled to the field operator ${\bf R}$.
It is seen that the field operator ${\bf R}$ is expressed in terms of the infinite bosonic field operators.
The noise operator in this approach are expressed in terms of these bosonic operators
at the initial time and the metric of the physical system.
It is stressed that the formalism tends to the normal Higgs model when the dissipative effects is neglected.

By using the perturbation theory, due to the dissipative effects, the
transition rate between energy levels is calculated. It is shown that
appreciable probabilities for transition are possible only if the
transition and reservoir's oscillators frequencies are nearly on
resonance, e.g., if, $E_{n}\simeq E_{m}-\hbar \omega$ (stimulated
emission) or by $E_{n}\simeq E_{m}+\hbar \omega$ (absorption).
Therefore, the particle on the sphere may be changed from
lower(upper) states to upper(lower) ones by selecting suitable
frequencies of the reservoir's oscillators.


\begin{thebibliography}{10}
%
\bibitem{1} P.W. Higgs, J. Phys. A: Math. Gen. 12 (1979) 309.
%
\bibitem{2} H.I. Leemon, J. Phys. A: Math. Gen. 12 (1979) 489.
%
\bibitem{4} A. Mahdifar , R. Roknizadeh  and M.H. Naderi, J. Phys. A: Math.
Gen. 39 (2006) 7003.
%
\bibitem{3} A.I. Solomon, Phys. Lett. A 196 (1994) 29; J. Katriel  and A.I. Solomon, Phys. Rev. A 49 (1994) 5149;
P. Shanta, S. Chaturvdi, V. Srinivasan
and R. Jagannathan, J. Phys. A: Math. Gen. 27 (1994) 6433.
%
\bibitem{5} A. Mahdifar, W. Vogel, T. Richter, R. Roknizadeh and M.H. Naderi, Phys. Rev. A 78 (2008) 063814.
%
\bibitem{6} A. Mahdifar, R. Roknizadeh and M.H. Naderi, Int. J. Geom.
Methods Mod. Phys. 9 (2012) 1250009.
%
\bibitem{6.5} A. Mahdifar, Int. J. Geom.
Methods Mod. Phys. 10 (2013) 1350028.
%
\bibitem{Mirza} A. Mahdifar, B. Mirza, R. Roknizadeh, J. Phys. A: Math.
Gen. 45(46) (2012) 465301.
%
\bibitem{Ford1965} G.W. Ford, M. Kac, and P. Mazur, J. Math. Phys. 6 (1965) 504;
G.W. Ford, J.T. Lewis, and R.F. O'Connell, Phys. Rev. A 37 (1988) 4419.
%
\bibitem{Mori1965} H. Mori, Progr. Theor. Phys. 33 (1965) 423.
%
\bibitem{Caldeira1983} A.O. Caldeira, A.J. Leggett, Physica A 121 (1983) 587; A.O. Caldeira, A.J. Leggett, Ann. Phys. 149 (1983) 374.
%
\bibitem{Leggett1987} A.J. Leggett, et al, Rev. Mod. Phys. 59 (1987) 1.
%
\bibitem{Schwinger1961} J. Schwinger, J. Math. Phys. 2 (1961) 407.
%
\bibitem{Keldysh1965} L.V. Keldysh, Sov. Phys. JETP 20 (1965) 1018.
%
\bibitem{Feynman1963} R. Feynman, F. Vernon, Ann. Phys. 24 (1963) 118.
%
\bibitem{Grabert1988} H. Grabert, P. Schramm, G.L. Ingold, Phys. Rep. 168 (1988) 115.
%
\bibitem{Hu1992} B.L. Hu, J.P. Paz, Y. Zhang, Phys. Rev. D 45,(1992) 2843.
%
\bibitem{Weiss2008} U. Weiss, Quantum Dissipative Systems, World Scientific
Publishing, Singapore, 2008.
%
\bibitem{Amooghorban2014} E. Amooghorban and F. Kheirandish, Int. J. Theor. Phys. 53 (2014) 2593.
%
\bibitem{Kheirandish2010} F. Kheirandish and E. Amooghorban, Phys. Rev. A 82 (2010) 042901;
E. Amooghorban, M. Wubs, N.A. Mortensen, and F. Kheirandish, Phys. Rev. A 84 (2011) 013806.
%
\bibitem{Amooshahi2010} M. Amooshahi, and E. Amooghorban, Ann. Phys. 325 (2010) 1976; F. Kheirandish, E. Amooghorban
 and M. Soltani, Phys. Rev. A 83 (2011) 032507;
%
\bibitem{Hopfield1958} J.J. Hopfield, Phys. Rev. 112 (1958) 1555.



\end{thebibliography}
\end{document}